# ParasNet: Fast Parasites Detection with Neural Networks

X. F. Xu, S. B. Lu, S. Talbot, T. Selvaraja

*Abstract*— Deep learning has dramatically improved the performance in many application areas such as image classification, object detection, speech recognition, drug discovery and etc since 2012. Where deep learning algorithms promise to discover the intricate hidden information inside the data by leveraging the large dataset, advanced model and computing power. Although deep learning techniques show medical expert level performance in a lot of medical applications, but some of the applications are still not explored or under explored due to the variation of the species. In this work, we studied the bright field based cell level Cryptosporidium and Giardia detection in the drink water with deep learning. Our experimental demonstrates that the new developed deep learning-based algorithm surpassed the handcrafted SVM based algorithm with above 97 percentage in accuracy and 700+fps in speed on embedded Jetson TX2 platform. Our research will lead to real-time and high accuracy label-free cell level Cryptosporidium and Giardia detection system in the future.

## I. Introduction

As the development of the machine learning technical, the deep learning based methods showed a breakthroughs performance in many areas such as image classification, object detection, semantic segmentation, decision making and etc [1][2][21][22][23][24]. Recently, the deep learning based methods also spread into medical imaging fields where It improved the performance a lots in computer-aided detection (CADe) and diagnosis (CADx), radionics, and medical image analysis [20].

Medical imaging analysis is an important task in biomedical where medical imaging techniques give the scientists the possibilities to visualize the systematic internal and external representation from organisms to cells. Overall, the scientists use ultrasound, radio wave(x-rays, laser) and magnetic(Magnetic Resonance Imaging) for generating medical images [3][4]. The digital representation of the medical images include medical ultrasonography, microscope based (Bright-field microscopy, Fluorescence microscopy, Confocal scanning microscopy, Scanning electron microscopy and etc.), magnetic resonance imaging (T1, T2, magnetic resonance angiography, functional MRI and etc.), flow cytometry cell images [5], light scattering cell images [6] and etc. Although deep learning techniques show medical expert level performance in some of those medical images analysing [7][8], however some of the areas are still not explored or under explored. The examples include cell level scattering images detection on Cryptosporidium and Giardia [9].

Cryptosporidium and Giardia are the cell level parasites which are wildly exist in contaminated drinking water, where contaminated drinking water is the important source spreading diarrhoea, cholera, dysentery, typhoid, and polio diseases. They even still can live after pass-through various chemical and filtration processes used in the water treatment. Some people can be very sick by infecting with those two type of parasites. World Health Organization (WHO) reports [10][11] that It estimated to cause 502,000 people deaths in each year [12].

The detail of label-free scattering imaging on Cryptosporidium, Giardia and other particles can be found in Figure 1. From the figure, we can see that the Cryptosporidium has a big grey band and followed by a black area. The Giardia has oval shape and more fringes than Cryptosporidium. Furthermore, the distance between consecutive fringes of Giardia is smaller.

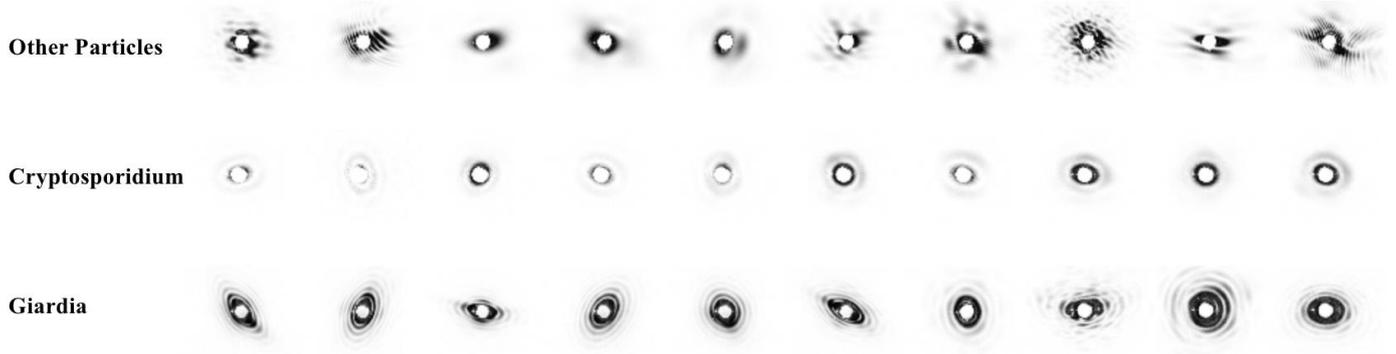

Fig. 1. Holograph images of cryptosporidium, giardia and other particles

In this work, we made the following contributions:

We built a cell level Cryptosporidium and Giardia scatter image dataset for Cryptosporidium and Giardia detection in the drink water.

We investigated the traditional feature engineering approaches for detecting the Cryptosporidium and Giardia.

We presented the ParasNet, a 8 layers convolutional neural network, which inputs a cell level scattering image of the particles inside the drinking water and outputs the probability of the parasites: Cryptosporidium and Giardia, for water quality inspection.

We optimized the ParasNet and run the algorithm on the embedded Jetson TX2 platform and archived up to 100fps, which enables the real-time parasites detection in the future.

## II. IMPLEMENTATION

### A. Dataset

In order to evaluate the performance of the different algorithms for the cell level Cryptosporidium and Giardia detection, we collected a lot images during the real experiments. We split several collected image sequences into training data collection and others into testing data collection. Inside those training data collection, we manually selected 5000 images each for Cryptosporidium, Giardia and other particles. We also random selected 1000 images each for building the testing dataset. Every image inside the Cryptosporidium and Giardia dataset has the dimension with 648 x 488 pixels and greyscale. The details of scattering imaging on Cryptosporidium, Giardia and other particles can be found in Figure 1. The Cryptosporidium has a big grey band and followed by a black area. The Giardia has oval shape and more fringes than Cryptosporidium. In addition, the distance between consecutive fringes is smaller. Other particles has a different noise background. Because the real Cryptosporidium and Giardia has distinct internal cell structure, size, and the orientation or position related to the MEMS microchannel, the scattering image shows complex pattern in the size, translation, illumination, deformation, intra-class variance and etc. For example, the big grey band of the Cryptosporidium is ellipse in some time but some time is cycle. The density of the big black band is also flickering. The fringes of the Giardia also show complex orientations. In term of the data selection, we ensured that the selected data contains significate variances in object size, pose, orientation and position.

### B. SIFT + SVM based algorithm

First of all, we evaluated the dataset with a hand-crafted SIFT features [13] with SVM and Naïve Bayes based pipeline is shown in Figure 2. Firstly, the input image is processed by de-noise and image quality enhance algorithms for better SIFT features detection in the next stage. Consequently, The enhanced image is extracted by SIFT algorithm for the SIFT key points which invariants to scale and orientation.

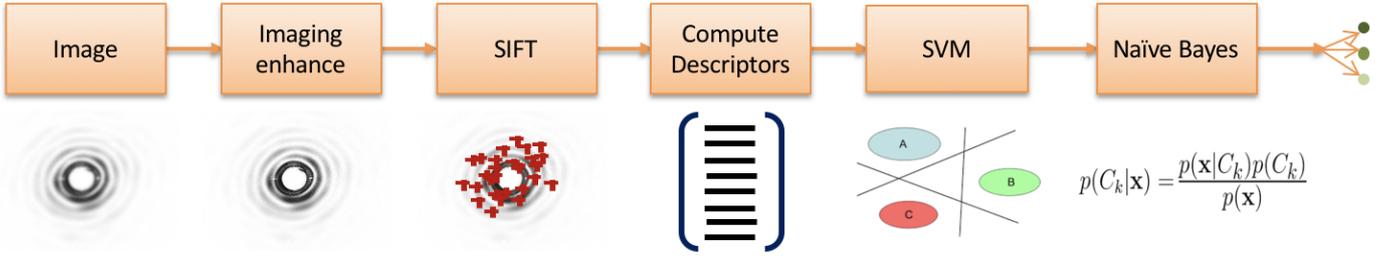

Fig. 2. SIFT+SVM based hand-crafted algorithm

The SIFT algorithm firstly finds the scale-space extrema in the Difference-of-Gaussian function convolved with the image as key points. Then excludes low contrast or strong edge responses. Finally assigns the orientation for determined key point by choosing the peak gradient magnitude of its neighbourhood points. Based on the SIFT key points, the SIFT descriptors are generated by calculating the points' gradient magnitude and orientations. Later, it builds up the gradient orientation histogram weighted by a Gaussian function and the gradient magnitude. Finally, they are passed through a trained score based SVM model [14][15]. Unlike the tradition linear SVM, the score based SVM can output the probabilities that indicts relationship between the SVM optimal hyperplane and the geometry distribution of the data in the feature space with how similar the test sample to all the categories. For the SVM based algorithm still has some problem in differentiating Cryptosporidium and Giardia, we also add one more Naïve Bayes classifier to improve the result.

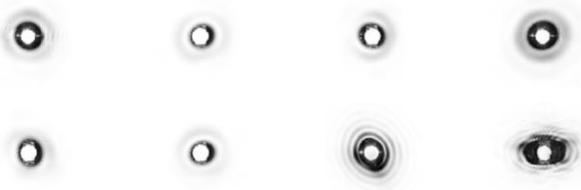

Fig. 3. Example of false detection images

In result, this SVM based algorithm can archive 84.5% accuracy on 1000 testing images for Cryptosporidium and 99.5% accuracy on 1000 testing images for Giardia. For the performance, it can only up to 5fps in processing speed on a Jetson TX2. The confusion matrix shows in Table 1 at below. From the result, we found that the algorithm cannot handle well while the images exist noise and the other particles are similar with Cryptosporidium and Giardia as well as the contrast of the imaging is very weak shows in Figure 6. It almost lacks in Cryptosporidium detection. There have 155 out of 1000 false detection in Cryptosporidium.

Table 1 Confusion Matrix

|  |  | PREDICTED | | |
|---|---|---|---|---|
|  |  | Others | Crypto. | Giardia |
| ACTUAL | Others | 1000 | 0 | 0 |
|  | Crypto. | 155 | 845 | 0 |
|  | Giardia | 5 | 0 | 995 |

C. *ParasNet*

In addition to evaluate the traditional hand-crafted based algorithms, we also design a new convolutional neural network based classification algorithm with latest ideas from the deep learning community.

In this approach, we formulated the parasites detection problem as a multiple classification problem, where the input is a scattering image X with 324x244 pixels (We down scale the original image with 648x488 pixels for faster computation speed) and the output is a vector $y$ (probability indictor on how much other particle, Cryptosporidium, or Giardia it is). The finally unweight binary cross entropy loss function is:

$$L(X,y) = \sum_{c=0}^{2}[-y_c \log p(Y_c = 1|X) - (1 - y_c)\log p(Y_c = 0|X)]$$

Where $p(Y_c = 1|X)$ indicts the probability of exist class c inside the input image X and $p(Y_c = 0|X)$ indicts the probability of class c does not inside the input image X.

*1) ParasNet Architecture*

ParasNet is an 8 layers deep neural network shows in Figure 4. It comprised 5 convolutional layers. Every convolutional layers combines with rectified linear unit (ReLU) non-linearity layer and pooling layer. Inside the convolutional layer, we use small receptive field filter 3x3 for convolution as suggest in [16]. The convolution is padded with zero pixel and stride is fixed to one pixel. In result, the spatial resolution of the output features layers are shrank 2 pixels after every convolution. A ReLU layer with f(x) = max (0, x) follows with the convolutional layer. Spatial pooling is performed by max-pooling with 2x2 pixels window and striding 2 pixels. Two Fully-Connected(FC) layers follow the whole CONV + ReLU + POOLING stack. The first FC layer has 128 output units and the second FC layer has 3 outputs units which indicts the class score of every classes: Other particle, Cryptosporidium and Giardia. The finally layer is the soft-max layer. It transforms the class score of every class to the classification probability of different classes.

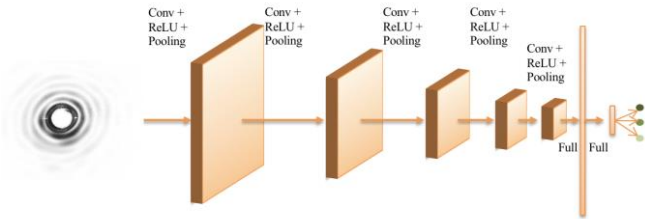

Fig. 4. The ParasNet

*2) Training*

We fed the labelled 15K images into the network with downscaled them to 324 x 244 pixels and normalized to [0, 1]. Furthermore, we also augmented them in real-time with random position transform in both direction, horizontal and vertical flipping, rotation and zoom. The weight are initialized by Glorot uniform initializer [17] and the network was trained by an end-to-end fashion with Adam stochastic optimizing algorithm [18]. The parameters for Adam are learning rate = 0.001, beta1 = 0.9, beta2 = 0.999. A learning rate decay also be used for training.

In order to find the best filter numbers of the ParasNet, we perform a parameters search over the whole dataset. Similar to learning curves, we got a curve with number of parameters in every layer vs the maximum test accuracy as shown in Figure 5.

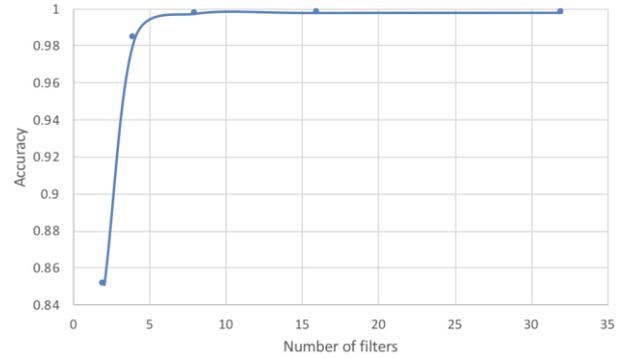

Fig. 5. Number of filters vs testing accuracy

From the Figure 5, we found that the test accuracy will be lesser than 98% when the filters number lesser than 4 filters per layer. As the filters number increases to 8, the testing accuracy of the ParasNet reaches the saturate point. Later, the accuracy increases lesser after this point. For less filters will be lesser computation, we selected 8 filters per layer as the best filter count parameter.

Finally, we got the detail configuration of ParasNet in the below Table 2. It includes 5 convolutional layers and 8 filters per convolutional layer. In total, it includes 43,891 parameters.

Table 2 The network configuration

| NAME | TYPE | OUTPUT SHAPE | PARAMS |
|---|---|---|---|
| INPUT IMAGES | | 242 x 322 x 8 | |
| CONV1 | Convolutional | 242 x 322 x 8 | 80 |
| POOLING1 | Max Pooling | 121 x 161 x 8 | 0 |
| CONV2 | Convolutional | 119 x 159 x 8 | 584 |
| POOLING2 | Max Pooling | 59 x 79 x 8 | 0 |
| CONV3 | Convolutional | 57 x 77 x 8 | 584 |
| POOLING3 | Max Pooling | 28 x 38 x 8 | 0 |
| CONV4 | Convolutional | 26 x 36 x 8 | 584 |
| POOLING4 | Max Pooling | 13 x 18 x 8 | 0 |
| CONV5 | Convolutional | 11 x 16 x 8 | 584 |
| POOLING5 | Max Pooling | 5 x 8 x 8 | 0 |
| DENSE1 | Fully connected | 128 | 41088 |
| DROPOUT | Dropout | 128 | 0 |
| DENSE2 | Fully connected | 3 | 387 |
| SOFTMAX | Softmax | 3 | 0 |

*3) Testing*

We evaluated the trained model on 3000 testing images. It is given trained ConvNet model and the input images with the batch size of 32 images. The images are rescaled to smaller network input size. Then the input batch is scaled down to [0 , 1] and put into the first layer. The class score map with the shape of batch size × number of classes are generated after one pass. Finally, we got 95.6% test accuracy over the 1000 testing images for Cryptosporidium and 99.5% test accuracy over the 1000 testing images for Giardia as shown in Table 3. The false detection images are shown in Figure 6.

Table 3 Confusion Matrix

| | | PREDICTED | | |
|---|---|---|---|---|
| | | Others | Crypto. | Giardia |
| ACTUAL | Others | 1000 | 0 | 0 |
| | Crypto. | 44 | 956 | 0 |
| | Giardia | 5 | 0 | 995 |

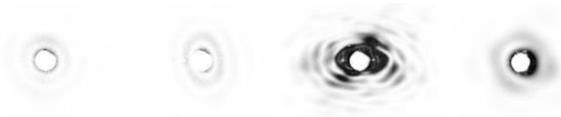

Fig. 6. Example of false detection images

*D. Peformance Comparsion*

We assessed the performance with SVM based algorithm (84.5% in Cryptosporidium testing accuracy). We found that the SIFT + SVM based algorithm is lack in Cryptosporidium detection with 155 out of 1000 false detection. But for the ParasNet, it has lesser false detection. From speed view, the SVM based algorithm is 20 times more slower than ParasNet on Jetson TX2 for it used hand-crafted features and complex pipeline. Furthermore, we met a lot of difficulty in turning the parameters of SIFT and SVM, but the ParasNet can do end-to-end based training.

Later, we examined the features generated by the last layer of ParasNet with t-distributed Stochastic Neighbour Embedding (t-SNE) [19] algorithm. As shown in Figure 7, the point in red, green or blue is a two dimensions projection from 128 dimensional vector of last CNN layer, the three classes are well separated in the point cloud with the Cryptosporidium in the bottom left, the Giardia in the top and others in the bottom right.

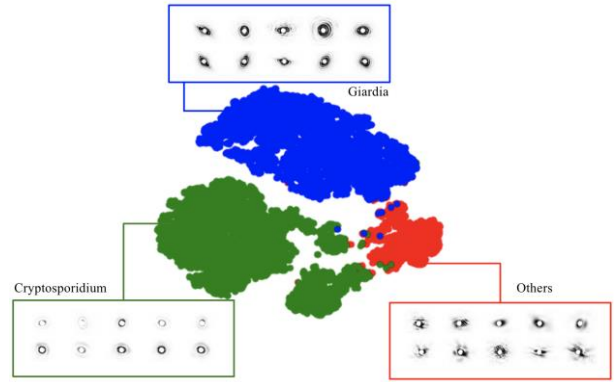

Fig. 7. The t-SNE visulization of the last hidden layer

Finally, we also evaluated the algorithm on Nvidia's Jetson TX2 platform for the performance test. It archived 100 images per second in classification test. Overall, the ParasNet surpass the SVM both in speed and accuracy. It can make a big progress in designing next generation cell level Cryptosporidium and Giardia detection system.

Table 4 Performance compasion

| | CRYPTO. ACCURACY | GIARDIA ACCURACY | SPEED |
|---|---|---|---|
| SVM BASED | 84.5% | 99.5% | ~5fps |
| CNN BASED | 95.6% | 99.5% | ~100fps |

## III. CONCLUSION

Drinking water's quality is crucial to human life while early detection the Cryptosporidium and Giardia can prevent the public in health dangerous conditions. Here we built a cell level Cryptosporidium and Giardia dataset and demonstrated the effectiveness of the deep convolutional neural network based — ParasNet together a traditional hand-crafted detection algorithm on parasites detection. The result shows that the deep learning based ParasNet can detect the Cryptosporidium and Giardia in the cell level scattering images with above 95.6% accuracy and run up to 100fps on embedded Jetson TX2 device. Our research can lead to product a lower cost and real-time cell level Cryptosporidium and Giardia detection system in the future. That will greatly improve human's life quality.